\documentclass[%
 reprint,
 amsmath,amssymb,
 aps,
]{revtex4-1}

\usepackage{tikz}
\usepackage{todonotes}
\usepackage[FIGTOPCAP,raggedright,hang]{subfigure}
\usepackage{graphicx}
\usepackage{dcolumn}
\usepackage{bm}
\usepackage{color,soul,url}

\begin{document}

\preprint{APS/123-QED}

\title{Bulk soliton dynamics in bosonic topological insulators}

\author{Jeremy L. Marzuola$^1$, Mikael C. Rechtsman$^2$, Braxton Osting$^3$, Miguel A. Bandres$^4$}
\affiliation{$^1$Department of Mathematics, University of North Carolina at Chapel Hill, NC 27599,\\ $^2$Department of Physics, The Pennsylvania State University, University Park, PA 16802\\  $^3$Department of Mathematics, University of Utah, Salt Lake City, UT 84112\\  $^4$CREOL, The College of Optics and Photonics, University of Central Florida, Orlando, Florida 32816, USA}
\date{\today}

\begin{abstract}
We theoretically explore the dynamics of spatial solitons in nonlinear/interacting bosonic topological insulators.  We employ a time-reversal broken Lieb-lattice analog of a Chern insulator and find that in the presence of a saturable nonlinearity, solitons bifurcate from a band of non-zero Chern number into the topological band gap with vortex-like structure on a sublattice.  We numerically demonstrate the existence stable vortex solitons for a range of parameters and that the lattice soliton dynamics are subject to the anomalous velocity associated with large Berry curvature at the topological Lieb band edge.  The features of the vortex solitons are well described by a new underlying continuum Dirac model.  We further show a new kind of interaction: when these topological solitons `bounce' off the edge of a finite structure, they create chiral edge states, and this give rise to an "anomalous" reflection of the soliton from the boundary.  
\end{abstract}

\pacs{Valid PACS appear here}
\maketitle

The insight by Haldane and Raghu \cite{haldane2008possible} that the topological properties of the quantum Hall effect were not limited to electrons propagating in solid-state materials, but could be applied to photons propagating in a photonic crystal, has led to a cross-disciplinary effort to explore and exploit topological robustness.  For instance, topological interface states that are highly robust to disorder have been observed in microwave photonic \cite{wang2009observation} and optical \cite{rechtsman2013photonic, hafezi2013imaging} platforms, in mechanical systems \cite{nash2015topological, susstrunk2015observation} and even in amorphous and quasicrystals lattices \cite{Irvine,bandres2016topological}.  In both ultracold atomic and photonic systems, topological invariants have been directly measured using bulk response \cite{rudner2009topological, jotzu2014experimental,atala2013direct, aidelsburger2015measuring, zeuner2015observation}.

The large emphasis of recent work has been demonstrating the presence of topological phenomena, such as protected edge states, in the linear (i.e., non-interacting) regime. 
A major question in the field has been whether nonlinear optical response in topological systems give rise to novel phenomena where topological character and nonlinearity are both essential.  Initial results on nonlinear topological systems include the prediction of bulk topological solitons \cite{lumer2013self}, soliton edge states \cite{ablowitz2014linear}, self-induced topological edge states \cite{katan2016induction,leykam2016edge}, and nonlinear effects in coupled-ring resonator systems \cite{leykam2018reconfigurable}.  

Here, we theoretically and numerically demonstrate and study the dynamics of solitons propagating in the bulk of a bosonic topological insulator and whose energies lies inside the topological band gap.  Our description applies equally well to a range of nonlinear wave systems, including optical waveguide arrays, planar photonic crystals, mechanical lattices, ultracold atomic systems in optical lattices, and others. The mechanism for the formation of these solitons is fundamentally topological: the wavefunction describing the soliton has $2\pi$ winding that is inherited from the Berry curvature of the topological band edge.  The inner core of the soliton corresponds to the `conducting bulk' of a material, and the exterior of the soliton behaves similarly to a topological edge state.  While previous work has demonstrated the existence of such solitons \cite{lumer2013self}, the dynamics thereof were not explored, and they were not linearly stable (though they did exhibit long lifetimes under some conditions).  In the present work, we demonstrate a model for which the topological solitons can be stable; derive a continuum model describing the solitons; demonstrate that the soliton dynamics include the anomalous velocity term arising from the non-zero Berry curvature of the band; and study the dynamics of the soliton interacting with the edge of the system.

For concreteness, we describe the system using the language of photonic waveguide arrays, where each site comprises a single-mode waveguide that is coupled in an evanescent manner to its neighbors in the lattice.    
\begin{figure*}[t]

\includegraphics[width=1\linewidth]{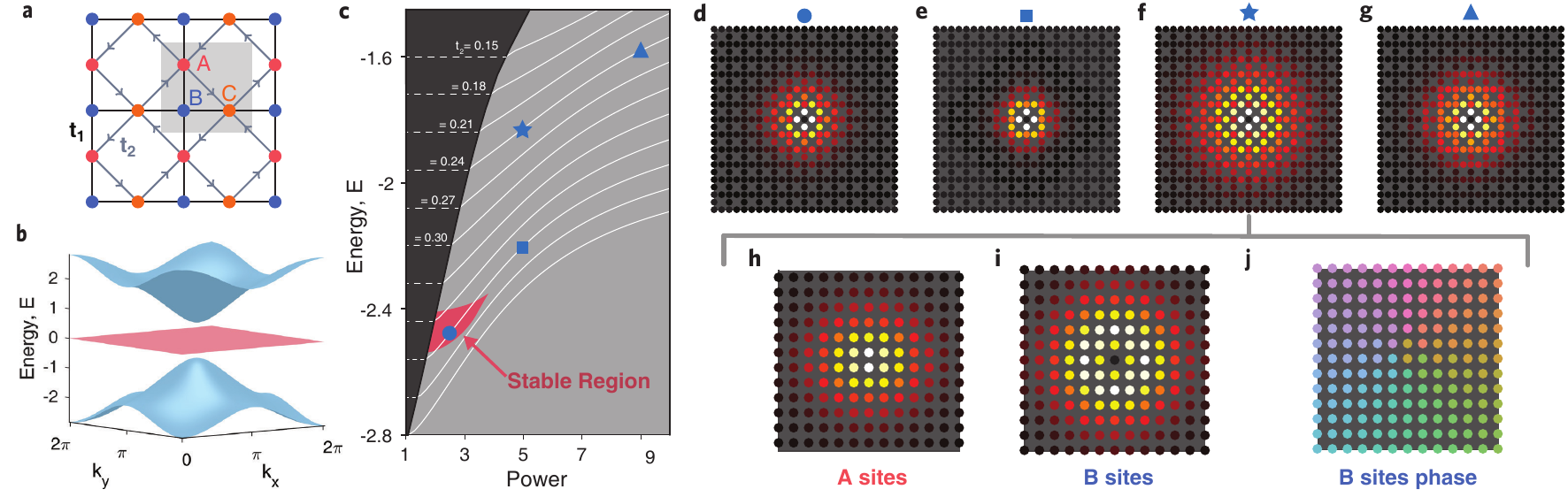} \\
  
\caption{\label{fig1} {\bf (a)} The topological Lieb lattice is composed of three sublattices, labelled {\color{red} A} (red), {\color{blue} B} (blue), and {\color{orange} C} (orange).
{\bf(b)} The three bands of the tight-binding Hamiltonian are plotted within the first Brillouin zone. {\bf (c)} Phase diagram of soliton stability, as a function of energy and power, {\bf (d-g)} Soliton wavefunction magnitude profiles for points indicated in (c). Note that the magnitude of the A and C sites is the same though they are out of phase by $\pi/2$ . {\bf(h)} Wavefunction magnitude profile at the A sites, {\bf(i)}  at the B sites; and {\bf(j)} the phase at the B sites (for the soliton in (f)).  Soliton parameters for (h-j) are: $s=1.0$, $t_1 = 1.0$, $t_2=0.4$, $P = 3.31$, $E=-2.4044$. 
}
\end{figure*}
To describe the dynamics of the light propagating in the nonlinear lattice we use the discrete nonlinear Schr\"odinger equation with saturable nonlinearity:
\begin{equation}
i\partial_z \psi_m = \sum_n H_{mn}\psi_n + \frac{|\psi_m|^2}{1+s|\psi_m|^2}\psi_m,\label{nls}
\end{equation}
where $\psi_m$ is the amplitude on the $m^{th}$ site (i.e., waveguide) of the lattice; $z$ is the distance of propagation along the waveguide axis (and acts as a temporal coordinate); $s$ is a parameter that defines the degree of saturation of the nonlinearity ($s=0$ defines a Kerr nonlinearity); and $H_{mn}$ is the linear Hamiltonian that is given by a Chern insulator that is based on a Lieb lattice. See Fig.~\ref{fig1}(a) for a depiction of the lattice model. In Bloch wavevector space, $(k_x,k_y)$, the Hamiltonian is parameterized by the ratio of the next-nearest to nearest-neighbor hopping strengths $|t_2/t_1|$:
 \begin{tiny}
\begin{align}
\label{Hamiltonian_TopLieb}
& H(k_x,k_y) = \\
&  \begin{bmatrix}
0 & 2  \cos \left(  \frac{k_x}{2} \right) &  4 i \frac{ t_2}{t_1} \sin \left(  \frac{k_x}{2} \right) \sin \left(  \frac{k_y}{2} \right) \\
 2 \cos\left(  \frac{k_x}{2} \right) & 0 & 2  \cos \left(  \frac{k_y}{2} \right) \\
 -4 i \frac{ t_2}{t_1} \sin \left(  \frac{k_x}{2} \right) \sin \left(  \frac{k_y}{2} \right) &  2  \cos \left(  \frac{k_y}{2} \right) & 0 
\end{bmatrix}. \notag
\end{align}
\end{tiny}   

 The model corresponds to an array of {\it helical} waveguides arranged in a Lieb lattice.  The helicity breaks $z$-reversal symmetry by acting like a periodic temporal drive; under the influence of this drive, the Lieb lattice opens a topological band gap and thus exhibits the band structure of a Chern insulator (band structure shown in Fig. \ref{fig1}(b)).  We use the Magnus expansion \cite{bukov2015universal} to approximate the $z$-dependent system with a $z$-independent Hamiltonian (this approximation is valid in the high-drive-frequency limit).  In the simplest non-trivial case, the nearest-neighbor hopping is defined by a purely real $t_1$ (which we henceforth take to be 1) and the next-neighbor hopping, $\pm i t_2$, is imaginary, and has alternating signs as indicated in Fig.~\ref{fig1}(a), similarly to the Haldane model \cite{haldane1988model} with hopping phase $\pi/2$.  It is the second-neighbor hopping that is responsible for opening up the topological gap of size $4t_2$.  In this model, the bottom and top bands have Chern numbers $-1$ and $+1$ respectively, and the flat band between them has Chern number $0$.

The solitons that we search for are solutions to the $z$-independent nonlinear Schr\"odinger equation of Eq. \ref{nls}, i.e., static solutions to that equation.
We use a self-consistent iteration procedure to solve for solitons localized entirely in the bulk (we use periodic boundary conditions), bifurcating from the bottom band.  We will classify the solitons by the parameters given by the nonlinear eigenvalue $E$ in the band gap as well as the power $P\equiv \langle \psi | \psi \rangle$.  A phase diagram showing soliton existence curves is depicted in Fig. \ref{fig1}(c) for different values of $t_2$.  We observe that the effective support of the solitons are related both to $P$ and $t_2$. In Fig. \ref{fig1}(d-g), the magnitude of the wavefunction is plotted for a range of powers indicated in Fig. \ref{fig1}(c) displaying the vortex like structure inherited from the topological band.  We note that beyond the work \cite{lumer2013self} in the Floquet topological setting, vortex-like solitons have been studied in many previous works, e.g., Refs. \cite{malomed2001discrete,kevrekidis2001bound}.  However, the vortex solitons that we found here arise from a different mechanism, namely bifurcation from a band of high Berry curvature.

 The magnitude and phase profile of a typical soliton from such a family is shown in Fig. \ref{fig1}(h-j) - note the $2\pi$ winding associated with the phase of the soliton on the $B$-sites.  This is shown clearly in Fig.~\ref{fig1}(j), which depicts the phase of the wavefunction with the plane wave component of the Bloch wave at $(k_x,k_y)=(\pi,\pi)$ factored out.  Being in a topological band gap, the mechanism for the formation of these solitons is different than previously-explored gap solitons \cite{christodoulides1988discrete, eggleton1996bragg, stegeman1999optical, kivshar2003optical, fleischer2003observation}.  The inner core of the soliton can be interpreted as metal-like in the sense that the potential generated by the soliton detunes the lattice such that at the soliton energy, the core of the soliton is formed from bulk bands. However, on the periphery of the soliton there is decay associated with the fact that the soliton energy lies within the bulk band gap, much like an edge state.

Unlike the previous solitons found in topological gaps \cite{lumer2013self}, the soliton we find here using saturable nonlinearity in the Lieb lattice can be linearly stable (in a certain parameter regime, see Fig.~\ref{fig1}(c)).  In other words, when small perturbations are added to it, it remains intact indefinitely.  This  is  a  crucial  ingredient  because it  allows  for the long-time dynamics of the soliton  to  be  probed  as  it  propagates  through  the  lattice.  In order to characterize its stability, we linearize Eq. (\ref{nls}) around the soliton wavefunction $\phi_0$.  Specifically, we consider perturbations around $\phi_0$ of the form: $\phi = e^{-iEz}(\phi_0 + \epsilon\delta\phi)$, where $\epsilon \delta \phi$ is the perturbation (with $\epsilon \ll 1$).  By diagonalizing the resulting equation for $\delta\phi$, we are able to diagnose whether perturbations around $\phi_0$ are purely oscillatory or exponentially growing, implying instability.  In Fig. \ref{fig1}(c), we plot soliton stability curves for a range of next-neighbor hopping, $t_2$ from $0.15$ to $0.45$ (and for fixed saturable nonlinearity parameter $s=1.0$).  In the black region, there is insufficient power for the solitons to localize (lattice solitons are known to have a non-zero power threshold in two dimensions \cite{weinstein1999excitation}).  Solitons in the gray region are such that the largest real part of all stability eigenvalues is greater than $10^{-14}$  (a non-zero real part indicates exponential growth, therefore instability).  In the stable (red) region, the real part of the instability eigenvalue reaches at most $10^{-14}$, indicating stability.

\begin{figure}[t]
\begin{center}
\includegraphics[width=8cm]
{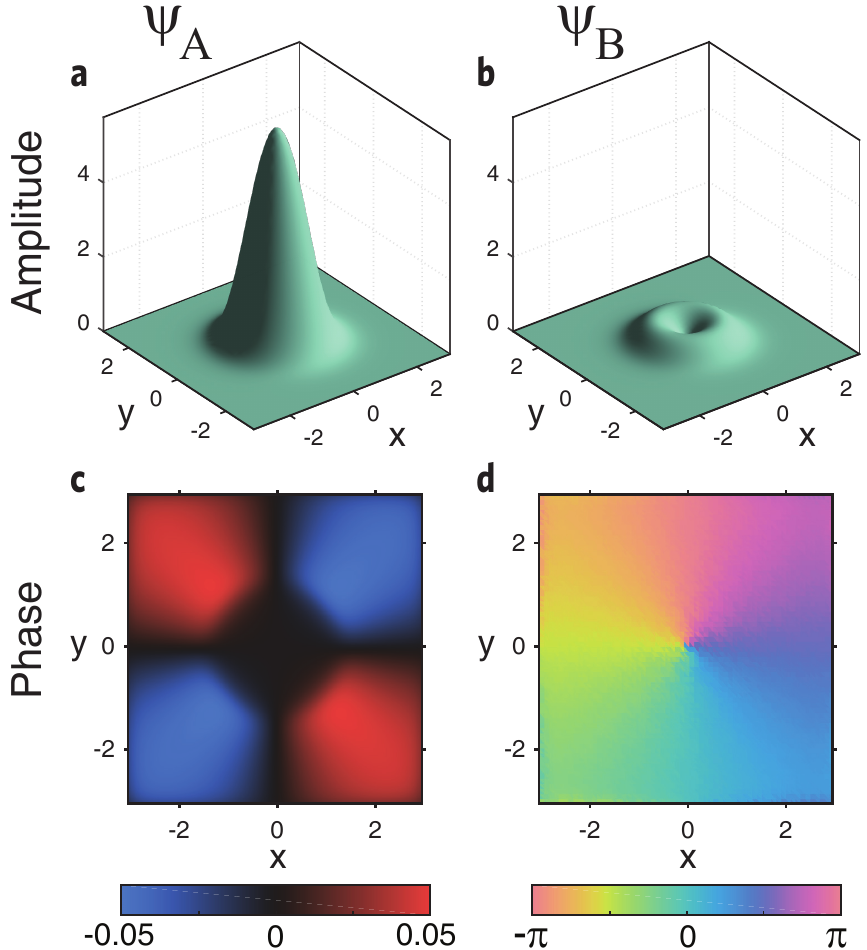}
\caption{\label{fig3} {\bf (a)}  The amplitude of the continuum soliton component corresponding to $\psi_A$ of the form  $\psi(r) + e^{2 i \theta} \eta(r)$, which is the same as that of $\psi_C$ of the form $-i (\psi(r) -e^{2 i \theta} \eta(r))  $, for some radial functions $\psi$ and $\eta$;
{\bf(b)} The continuum soliton component corresponding to $\psi_B$;
{\bf(c)} The continuum soliton phase component corresponding to $\psi_A$, which is a $\pi/2$ rotation of that from $\psi_C$;
{\bf(d)} The continuum soliton phase component corresponding to $\psi_B$.}
\end{center}
\end{figure}

Using the effective mass (i.e., two-scale) expansion, we are able to derive a continuum description of the soliton in the vicinity of the Dirac point.  To do this, we expand around the Brillouin zone corner (i.e., wavevector ${\bf k}=(\pi, \pi)$-point) - see Fig. \ref{fig1}(b).  Since the Lieb lattice has a three-member basis, we obtain three coupled nonlinear partial differential equations that describe the soliton, namely:
\begin{equation}
 H_{cont} \vec{\psi}+ \frac{|\vec{\psi}|^2}{1+s|\vec{\psi}|^2} \vec{\psi} = E \vec{\psi}, 
\label{continuum}
\end{equation}
with 
\begin{small}
\begin{equation}
H_{cont} = \left[  \begin{array}{ccc}
0 &  i   \partial_x & 4 i \frac{\tau_{2}}{\tau_{1}}  \\
 i   \partial_x  & 0 &  i   \partial_y \\
-4 i \frac{\tau_{2}}{\tau_{1}}   &  i  \partial_y & 0 
\end{array} \right],
\label{topLieb_TB}
\end{equation}
\end{small}
where $\tau_{2}/\tau_1$ is the continuum equivalent of $t_2/t_1$ above. Indeed, in the derivation, we have that $\tau_{2}/\tau_1 = t_2/t_1$, however it can be useful to scale the continuum parameters for computational purposes so we allow the continuum coefficients to be more general.   In terms of the three-dimensional Gell-Mann matrices $\lambda_i$, we can write the system
\begin{small}
\begin{equation}
H_{cont} = ( i \partial_x \lambda_1  + i \partial_y \lambda_6) - \frac{\tau_{2}}{\tau_1} \lambda_5.
\label{topLieb_TB_GM}
\end{equation}
\end{small}
These represent a novel nonlinear Dirac-type system, which is related to those discussed in for instance \cite{merle1988existence,esteban1995stationary,esteban1996stationary,esteban2002overview,boussaid2017nonrelativistic}.  We may again solve these equations using self-consistent iteration, and we obtain a branch of solitons.  We solve for soliton solutions of Eq. \eqref{continuum} and plot the results (for $\tau_1 = 1.0$, $\tau_{2} = 5.0$, $P = 10.0$) in Fig. \ref{fig3}.  The amplitude and phase profiles of the resulting soliton are shown in Fig. \ref{fig3}(a-d). They clearly match with those obtained for the lattice, as shown in Fig. \ref{fig1}(h-j): in particular, the same winding of the phase profile is observed.  

\begin{figure}[t]
\subfigure[\hspace{3.5cm}]{
\includegraphics[width=0.48\linewidth]{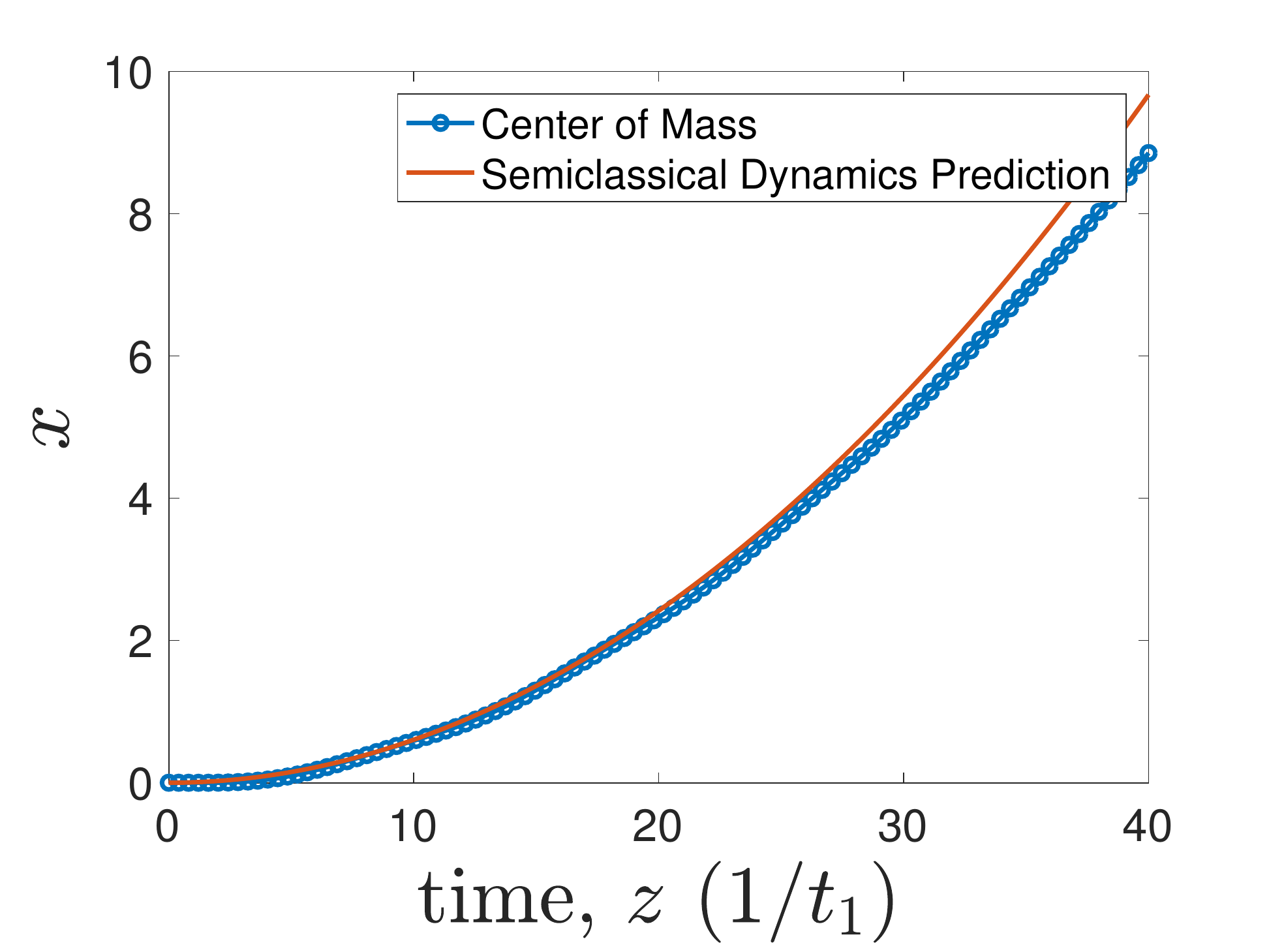}}
\subfigure[\hspace{3.5cm}]{
\includegraphics[width=0.48\linewidth]{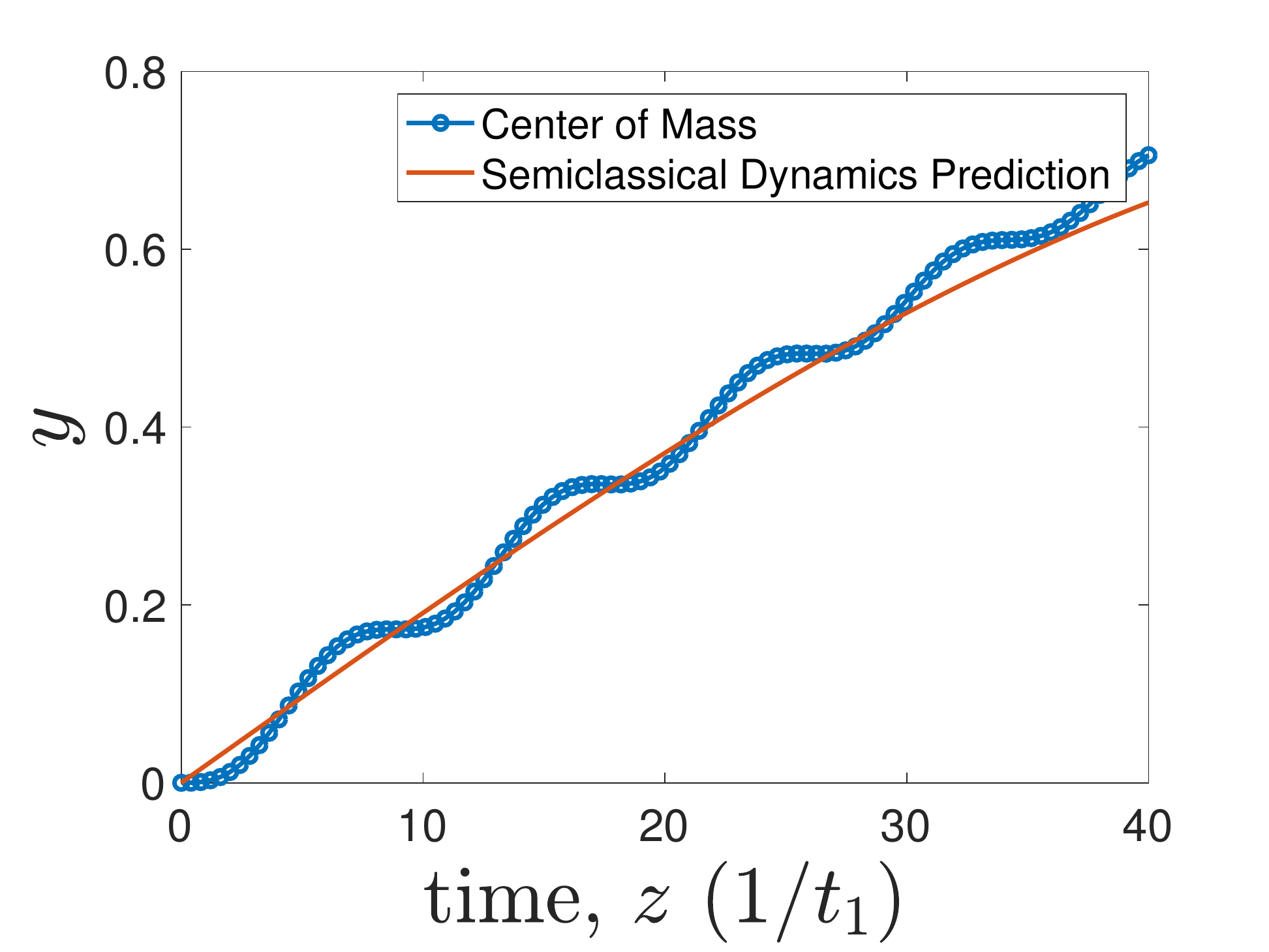}}
 \\
 \subfigure[\hspace{3.5cm}]{
\includegraphics[width=0.46\linewidth]{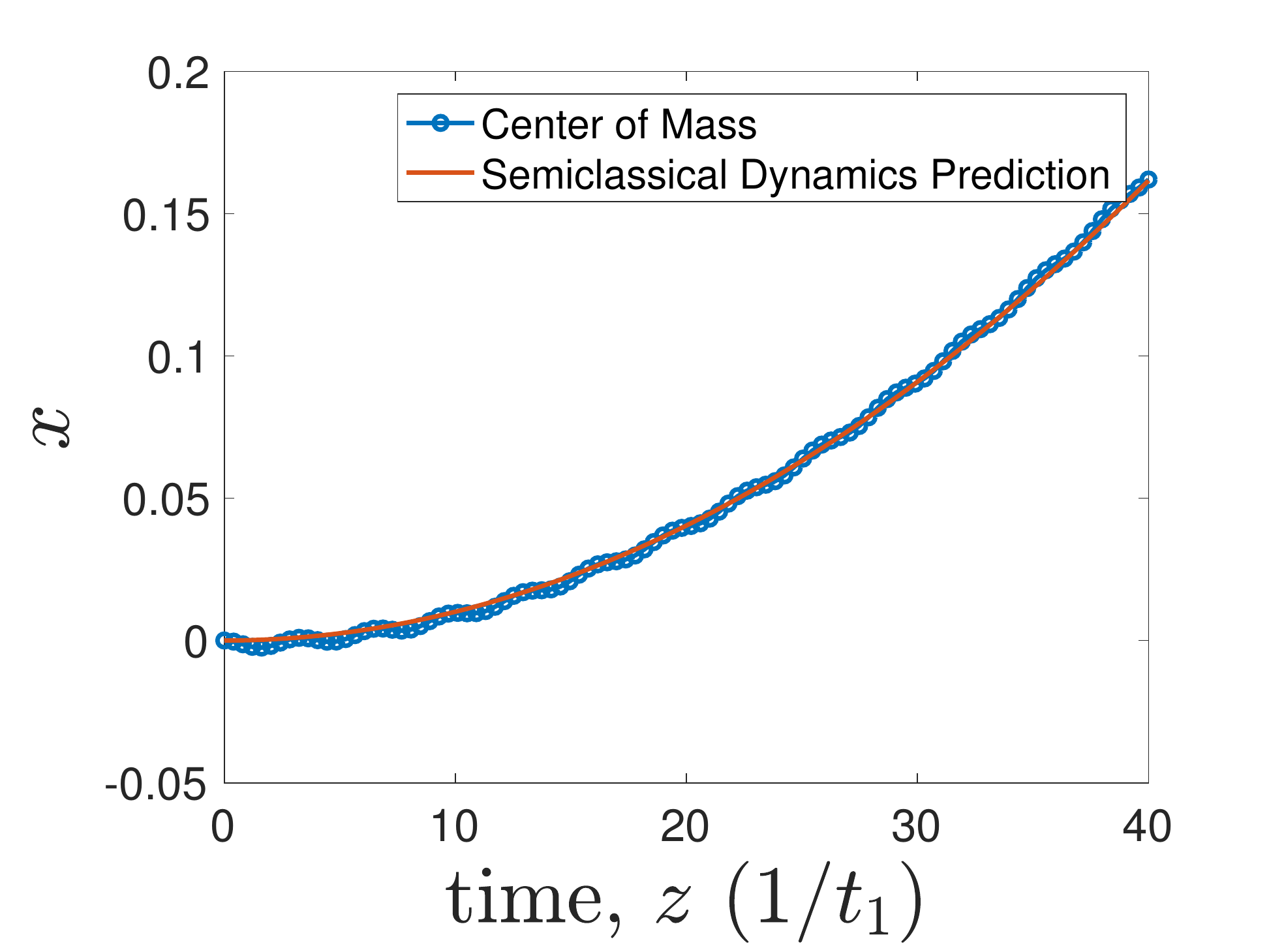}}
\subfigure[\hspace{3.5cm}]{
\includegraphics[width=0.46\linewidth]{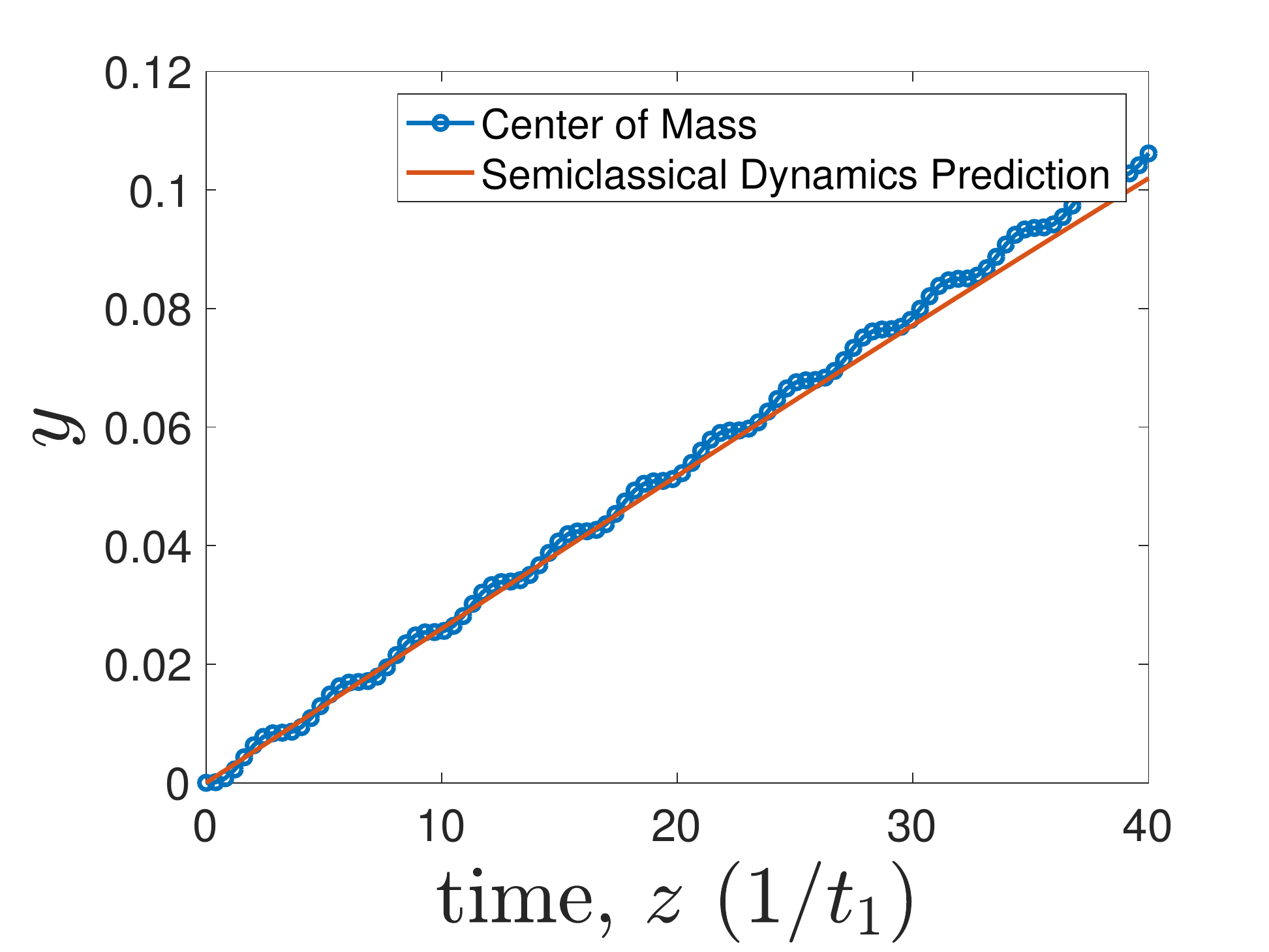}}
\caption{ Evolution of the center of mass, $[x(z),y(z)]$ (blue lines), of a low power topological soliton with (a,b) $t_2 = 0.18$, (c,d) $t_2 = 0.49$ . The red line depicted the expected semiclassical dynamics trajectory. Notice the ZB oscillations that are due the population of other bands.}
\label{fig4}
\end{figure}

The dynamics of wavepackets form from bands with non-zero Berry curvature is non-trivial: the semiclassical equations that describe its evolution under a external forces include an anonalous velocity term \cite{karplus1954hall}.  Our topological solitons live inside the band gap, and therefore do not belong to any band. And interesting question is what governs the dynamics of these topological solitons. We found that for weak nonlinearities and short times the topological soliton behaves as a wavepacket created from the band it bifurcates, and since topological solitons bifurcate point with large Berry curvature, they will have an anomalous velocity. 
Now, we return to the discrete model to show the effect of the anomalous velocity term on the soliton dynamics, making the assumption that the soliton is a wavepacket that largely spectrally occupies the bottom band (from which it bifurcates).  The semiclassical equations describing the center-of-mass position, ${\bf r}$, and velocity, ${\bf v}$ are:
\begin{equation}
\partial_z {\bf k} = {\bf F}; \ \ \partial_z {\bf r} = \nabla_{\bf k} \epsilon({\bf k}) - {\bf F} \times \Omega({\bf k}),
\label{semiclassical}
\end{equation}
where ${\bf F}$ is the force (i.e., potential gradient) applied to the system; $\epsilon({\bf k})$ represents the linear energy band structure, and $\Omega({\bf k})$ is the Berry curvature at Bloch wavevector ${\bf k}=(k_x,k_y)$.  In order to demonstrate that the soliton obeys these dynamics, we take one soliton with parameters $s=1.0$, $t_1=1.0$, $t_2 = .18$, $P=4.1202$, $E=-1.7037$ and one with $s=1.0$, $t_1=1.0$, $t_2 = .49$, $P=0.5020$, $E=-2.9598$. For these parameters, the solitons are unstable on time scales much longer than the dynamics we describe below.  We introduce an electric field in the $x$-direction (i.e., a potential gradient of ${\bf F}=1/100 \hat{x}$ per lattice spacing) and evolve the soliton in $z$ according to Eq. (\ref{nls}) using the fourth-order Runge-Kutta method. The evolution of the center of mass, $(x(z),y(z))$, of the soliton is shown in Fig. \ref{fig4}.   The time scale of the simulation is such that the soliton travels roughly $10$ unit cells for $t_2 = .18$.  Interestingly, we found that in the presence of the uniform electric field, it is possible to analytically solve (i.e., integrate) Eqs. \eqref{semiclassical}.  First, making the effective mass expansion in the vicinity of the Brillouin zone vertex of the Lieb lattice, from which the soliton bifurcates, we find a Berry curvature of: $\Omega = (\alpha |{\bf k}|^2+\beta)$, where ${\bf k}$ is the deviation in wave vector from the Brillouin zone vertex, $\alpha=3/512t_2^4$, and $\beta=-1/16t_2^2$.  By linearizing the dispersion of the bottom band, we obtain the wavepacket displacement:
\begin{equation}
{\bf r}(z) = (\mu F z^2, \alpha F^3 z^3/3 + \beta F z),
\end{equation}
where $\mu=t_2/2-1/8t_2$ is the curvature of the lower band.  
When we compare this analytic approximation to the numerical evolution of the topological soliton we observed a clear agreement, as depicted in Fig. \ref{fig4}.
 The oscillations in the $y$-component of the center of mass arise from {\it $Zitterbewegung$} (ZB) due to partial population of the other bands. An indication of these is the fact that the frequency of the oscillations is consistent with the ZB oscillations study in  \cite{dreisow2010classical}  for linear systems, i.e. for a spectral gap, $4 t_2$,
the frequency of ZB oscillations is $\omega = 4 t_2$. 

By increasing the power, on a comparable time scale the curves begin to diverge further from the semiclassical prediction as a result of nonlinear effects.

\begin{figure}[t]
\includegraphics[width=1\linewidth]{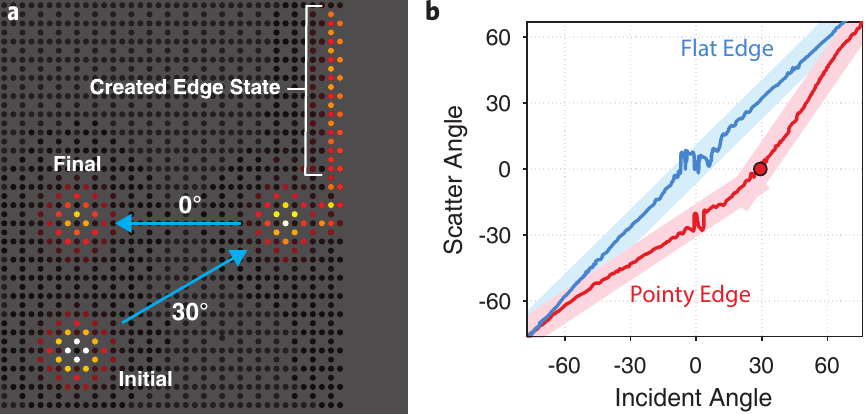}
\caption{ {\bf (a)} Sketch of the reflection of a soliton from the pointy edge of the lattice. The soliton travels with a $30^{\circ}$ angle with respect to the horizontal axis; when it reaches the pointy edge it generates a topological edge mode that takes all its momentum in the y-direction (in this case), in such a way that the soliton is reflected with a $0^{\circ}$ angle. The intensity of the topological edge mode is renormalized in order to clearly visualize it. {\bf (b)} Reflected angle as a function of the incident angle of a soliton colliding with the flat edge and the pointy edge of the lattice.  The reflection from the flat edge closely follows the expected behavior where the reflected angle is equal to the incident angle. However, the reflection from the pointy edge is anomalous because the soliton strongly populates a topological edge mode, which absorbs part of the y-momentum, when it scatters from the edge. The red circle represents the scattering event depicted in (a).}
\label{mig}
\end{figure}

We now describe an interesting new phenomenon that arises due to the topological nature of the solitons. Solitons in trivial systems reside spectrally within band gaps, where they are isolated from interaction with any other states of the system. For this reason, they behave like particles -- they move with constant velocity in the bulk of the system, and they perfectly bounce from the edge of the system, following the rules of momentum conservation parallel to the edge. However, topological solitons in finite systems are no longer isolated because topological edge states also reside in the topological band gap of the system. Thus, topological solitons can interact with and induce topological edge modes when they bounce from the edges of the system. To show this, we take a topological soliton (Fig.~\ref{fig1}(e)) and add a linear boost in momentum. We observe that when the solitons bounce from a `pointy' edge of the Lieb lattice (at right in Fig. \ref{mig}(a)), they populate a topological edge state, as shown in Fig. \ref{mig}(a) and Movie 1. This topological edge state takes away momentum from the soliton giving rise to an anomalous reflection angle for the soliton. This phenomenon is clearly shown in the plot of the reflected angle as a function of the incident angle (red line in Fig. \ref{mig}(b)).  We find that the soliton behaves differently when it bounces off the `flat' (at left of Fig. \ref{mig}(a)) and pointy edges, which have slow and fast group velocities, respectively.  In particular, the fast group velocity of the flat edge corresponds to low density of states (DOS), implying that the edge state is minimally populated while the soliton bounces.  Conversely, the pointy edge has low group velocity, thus relatively high DOS, and thus the edge state gets more populated (see Supplementary Movies 1 and 2).

To conclude, we have introduced a family of solitons that reside in a topological band gap that are vortex-like in character, owing to their bifurcation from a band of high Berry curvature.  The features of the vortex solitons are well described by a new underlying continuum Dirac model, and the dynamics of these solitons are impacted by the topological band structure from which they bifurcate.  These results give rise to a number of interesting directions, including the exploration of the nature of solitons in the bulk of higher Chern number lattices, soliton dynamics in the long-time regime (i.e., when they deviate from semiclassical behavior), and how such solitons interact with one another. This falls within a larger rubric of the exploration of interactions and topology in bosonic systems, and how novel behavior can emerge from the interplay of these two effects.

\subsection*{Acknowledgements}
J.L.M. was supported in part by NSF Applied Math Grant DMS-1312874 and NSF CAREER Grant DMS-1352353 and thanks Panos Kevrekidis, Jianfeng Lu and Daniel Spirn for helpful conversations about quantum vortices.  M.C.R. acknowledges support from the ONR-YIP program under grant number N00014-18-1-2595, the National Science Foundation (grant numbers ECCS-1509546 and DMS-1620422), the Packard Foundation (2017-66821) and the Kaufman Foundation (KA2017-91788).

\bibliography{refs}

\begin{thebibliography}{35}%
\makeatletter
\providecommand \@ifxundefined [1]{%
 \@ifx{#1\undefined}
}%
\providecommand \@ifnum [1]{%
 \ifnum #1\expandafter \@firstoftwo
 \else \expandafter \@secondoftwo
 \fi
}%
\providecommand \@ifx [1]{%
 \ifx #1\expandafter \@firstoftwo
 \else \expandafter \@secondoftwo
 \fi
}%
\providecommand \natexlab [1]{#1}%
\providecommand \enquote  [1]{``#1''}%
\providecommand \bibnamefont  [1]{#1}%
\providecommand \bibfnamefont [1]{#1}%
\providecommand \citenamefont [1]{#1}%
\providecommand \href@noop [0]{\@secondoftwo}%
\providecommand \href [0]{\begingroup \@sanitize@url \@href}%
\providecommand \@href[1]{\@@startlink{#1}\@@href}%
\providecommand \@@href[1]{\endgroup#1\@@endlink}%
\providecommand \@sanitize@url [0]{\catcode `\\12\catcode `\$12\catcode
  `\&12\catcode `\#12\catcode `\^12\catcode `\_12\catcode `\%12\relax}%
\providecommand \@@startlink[1]{}%
\providecommand \@@endlink[0]{}%
\providecommand \url  [0]{\begingroup\@sanitize@url \@url }%
\providecommand \@url [1]{\endgroup\@href {#1}{\urlprefix }}%
\providecommand \urlprefix  [0]{URL }%
\providecommand \Eprint [0]{\href }%
\providecommand \doibase [0]{http://dx.doi.org/}%
\providecommand \selectlanguage [0]{\@gobble}%
\providecommand \bibinfo  [0]{\@secondoftwo}%
\providecommand \bibfield  [0]{\@secondoftwo}%
\providecommand \translation [1]{[#1]}%
\providecommand \BibitemOpen [0]{}%
\providecommand \bibitemStop [0]{}%
\providecommand \bibitemNoStop [0]{.\EOS\space}%
\providecommand \EOS [0]{\spacefactor3000\relax}%
\providecommand \BibitemShut  [1]{\csname bibitem#1\endcsname}%
\let\auto@bib@innerbib\@empty
\bibitem [{\citenamefont {Haldane}\ and\ \citenamefont
  {Raghu}(2008)}]{haldane2008possible}%
  \BibitemOpen
  \bibfield  {author} {\bibinfo {author} {\bibfnamefont {F.}~\bibnamefont
  {Haldane}}\ and\ \bibinfo {author} {\bibfnamefont {S.}~\bibnamefont
  {Raghu}},\ }\href@noop {} {\bibfield  {journal} {\bibinfo  {journal}
  {{P}hysical {R}eview {L}etters}\ }\textbf {\bibinfo {volume} {100}},\
  \bibinfo {pages} {013904} (\bibinfo {year} {2008})}\BibitemShut {NoStop}%
\bibitem [{\citenamefont {Wang}\ \emph {et~al.}(2009)\citenamefont {Wang},
  \citenamefont {Chong}, \citenamefont {Joannopoulos},\ and\ \citenamefont
  {Solja{\v{c}}i{\'c}}}]{wang2009observation}%
  \BibitemOpen
  \bibfield  {author} {\bibinfo {author} {\bibfnamefont {Z.}~\bibnamefont
  {Wang}}, \bibinfo {author} {\bibfnamefont {Y.}~\bibnamefont {Chong}},
  \bibinfo {author} {\bibfnamefont {J.~D.}\ \bibnamefont {Joannopoulos}}, \
  and\ \bibinfo {author} {\bibfnamefont {M.}~\bibnamefont
  {Solja{\v{c}}i{\'c}}},\ }\href@noop {} {\bibfield  {journal} {\bibinfo
  {journal} {Nature}\ }\textbf {\bibinfo {volume} {461}},\ \bibinfo {pages}
  {772} (\bibinfo {year} {2009})}\BibitemShut {NoStop}%
\bibitem [{\citenamefont {Rechtsman}\ \emph {et~al.}(2013)\citenamefont
  {Rechtsman}, \citenamefont {Zeuner}, \citenamefont {Plotnik}, \citenamefont
  {Lumer}, \citenamefont {Podolsky}, \citenamefont {Dreisow}, \citenamefont
  {Nolte}, \citenamefont {Segev},\ and\ \citenamefont
  {Szameit}}]{rechtsman2013photonic}%
  \BibitemOpen
  \bibfield  {author} {\bibinfo {author} {\bibfnamefont {M.~C.}\ \bibnamefont
  {Rechtsman}}, \bibinfo {author} {\bibfnamefont {J.~M.}\ \bibnamefont
  {Zeuner}}, \bibinfo {author} {\bibfnamefont {Y.}~\bibnamefont {Plotnik}},
  \bibinfo {author} {\bibfnamefont {Y.}~\bibnamefont {Lumer}}, \bibinfo
  {author} {\bibfnamefont {D.}~\bibnamefont {Podolsky}}, \bibinfo {author}
  {\bibfnamefont {F.}~\bibnamefont {Dreisow}}, \bibinfo {author} {\bibfnamefont
  {S.}~\bibnamefont {Nolte}}, \bibinfo {author} {\bibfnamefont
  {M.}~\bibnamefont {Segev}}, \ and\ \bibinfo {author} {\bibfnamefont
  {A.}~\bibnamefont {Szameit}},\ }\href@noop {} {\bibfield  {journal} {\bibinfo
   {journal} {Nature}\ }\textbf {\bibinfo {volume} {496}},\ \bibinfo {pages}
  {196} (\bibinfo {year} {2013})}\BibitemShut {NoStop}%
\bibitem [{\citenamefont {Hafezi}\ \emph {et~al.}(2013)\citenamefont {Hafezi},
  \citenamefont {Mittal}, \citenamefont {Fan}, \citenamefont {Migdall},\ and\
  \citenamefont {Taylor}}]{hafezi2013imaging}%
  \BibitemOpen
  \bibfield  {author} {\bibinfo {author} {\bibfnamefont {M.}~\bibnamefont
  {Hafezi}}, \bibinfo {author} {\bibfnamefont {S.}~\bibnamefont {Mittal}},
  \bibinfo {author} {\bibfnamefont {J.}~\bibnamefont {Fan}}, \bibinfo {author}
  {\bibfnamefont {A.}~\bibnamefont {Migdall}}, \ and\ \bibinfo {author}
  {\bibfnamefont {J.}~\bibnamefont {Taylor}},\ }\href@noop {} {\bibfield
  {journal} {\bibinfo  {journal} {Nature Photonics}\ }\textbf {\bibinfo
  {volume} {7}},\ \bibinfo {pages} {1001} (\bibinfo {year} {2013})}\BibitemShut
  {NoStop}%
\bibitem [{\citenamefont {Nash}\ \emph {et~al.}(2015)\citenamefont {Nash},
  \citenamefont {Kleckner}, \citenamefont {Read}, \citenamefont {Vitelli},
  \citenamefont {Turner},\ and\ \citenamefont {Irvine}}]{nash2015topological}%
  \BibitemOpen
  \bibfield  {author} {\bibinfo {author} {\bibfnamefont {L.~M.}\ \bibnamefont
  {Nash}}, \bibinfo {author} {\bibfnamefont {D.}~\bibnamefont {Kleckner}},
  \bibinfo {author} {\bibfnamefont {A.}~\bibnamefont {Read}}, \bibinfo {author}
  {\bibfnamefont {V.}~\bibnamefont {Vitelli}}, \bibinfo {author} {\bibfnamefont
  {A.~M.}\ \bibnamefont {Turner}}, \ and\ \bibinfo {author} {\bibfnamefont
  {W.~T.}\ \bibnamefont {Irvine}},\ }\href@noop {} {\bibfield  {journal}
  {\bibinfo  {journal} {Proceedings of the National Academy of Sciences}\
  }\textbf {\bibinfo {volume} {112}},\ \bibinfo {pages} {14495} (\bibinfo
  {year} {2015})}\BibitemShut {NoStop}%
\bibitem [{\citenamefont {S{\"u}sstrunk}\ and\ \citenamefont
  {Huber}(2015)}]{susstrunk2015observation}%
  \BibitemOpen
  \bibfield  {author} {\bibinfo {author} {\bibfnamefont {R.}~\bibnamefont
  {S{\"u}sstrunk}}\ and\ \bibinfo {author} {\bibfnamefont {S.~D.}\ \bibnamefont
  {Huber}},\ }\href@noop {} {\bibfield  {journal} {\bibinfo  {journal}
  {Science}\ }\textbf {\bibinfo {volume} {349}},\ \bibinfo {pages} {47}
  (\bibinfo {year} {2015})}\BibitemShut {NoStop}%
\bibitem [{\citenamefont {Mitchell}\ \emph {et~al.}(2018)\citenamefont
  {Mitchell}, \citenamefont {Nash}, \citenamefont {Hexner}, \citenamefont
  {Turner},\ and\ \citenamefont {Irvine}}]{Irvine}%
  \BibitemOpen
  \bibfield  {author} {\bibinfo {author} {\bibfnamefont {N.~P.}\ \bibnamefont
  {Mitchell}}, \bibinfo {author} {\bibfnamefont {L.~M.}\ \bibnamefont {Nash}},
  \bibinfo {author} {\bibfnamefont {D.}~\bibnamefont {Hexner}}, \bibinfo
  {author} {\bibfnamefont {A.~M.}\ \bibnamefont {Turner}}, \ and\ \bibinfo
  {author} {\bibfnamefont {W.~T.~M.}\ \bibnamefont {Irvine}},\ }\href {\doibase
  10.1038/s41567-017-0024-5} {\bibfield  {journal} {\bibinfo  {journal} {Nature
  Physics}\ }\textbf {\bibinfo {volume} {14}},\ \bibinfo {pages} {380}
  (\bibinfo {year} {2018})}\BibitemShut {NoStop}%
\bibitem [{\citenamefont {Bandres}\ \emph {et~al.}(2016)\citenamefont
  {Bandres}, \citenamefont {Rechtsman},\ and\ \citenamefont
  {Segev}}]{bandres2016topological}%
  \BibitemOpen
  \bibfield  {author} {\bibinfo {author} {\bibfnamefont {M.~A.}\ \bibnamefont
  {Bandres}}, \bibinfo {author} {\bibfnamefont {M.~C.}\ \bibnamefont
  {Rechtsman}}, \ and\ \bibinfo {author} {\bibfnamefont {M.}~\bibnamefont
  {Segev}},\ }\href@noop {} {\bibfield  {journal} {\bibinfo  {journal}
  {Physical Review X}\ }\textbf {\bibinfo {volume} {6}},\ \bibinfo {pages}
  {011016} (\bibinfo {year} {2016})}\BibitemShut {NoStop}%
\bibitem [{\citenamefont {Rudner}\ and\ \citenamefont
  {Levitov}(2009)}]{rudner2009topological}%
  \BibitemOpen
  \bibfield  {author} {\bibinfo {author} {\bibfnamefont {M.~S.}\ \bibnamefont
  {Rudner}}\ and\ \bibinfo {author} {\bibfnamefont {L.}~\bibnamefont
  {Levitov}},\ }\href@noop {} {\bibfield  {journal} {\bibinfo  {journal}
  {Physical {R}eview {L}etters}\ }\textbf {\bibinfo {volume} {102}},\ \bibinfo
  {pages} {065703} (\bibinfo {year} {2009})}\BibitemShut {NoStop}%
\bibitem [{\citenamefont {Jotzu}\ \emph {et~al.}(2014)\citenamefont {Jotzu},
  \citenamefont {Messer}, \citenamefont {Desbuquois}, \citenamefont {Lebrat},
  \citenamefont {Uehlinger}, \citenamefont {Greif},\ and\ \citenamefont
  {Esslinger}}]{jotzu2014experimental}%
  \BibitemOpen
  \bibfield  {author} {\bibinfo {author} {\bibfnamefont {G.}~\bibnamefont
  {Jotzu}}, \bibinfo {author} {\bibfnamefont {M.}~\bibnamefont {Messer}},
  \bibinfo {author} {\bibfnamefont {R.}~\bibnamefont {Desbuquois}}, \bibinfo
  {author} {\bibfnamefont {M.}~\bibnamefont {Lebrat}}, \bibinfo {author}
  {\bibfnamefont {T.}~\bibnamefont {Uehlinger}}, \bibinfo {author}
  {\bibfnamefont {D.}~\bibnamefont {Greif}}, \ and\ \bibinfo {author}
  {\bibfnamefont {T.}~\bibnamefont {Esslinger}},\ }\href@noop {} {\bibfield
  {journal} {\bibinfo  {journal} {Nature}\ }\textbf {\bibinfo {volume} {515}},\
  \bibinfo {pages} {237} (\bibinfo {year} {2014})}\BibitemShut {NoStop}%
\bibitem [{\citenamefont {Atala}\ \emph {et~al.}(2013)\citenamefont {Atala},
  \citenamefont {Aidelsburger}, \citenamefont {Barreiro}, \citenamefont
  {Abanin}, \citenamefont {Kitagawa}, \citenamefont {Demler},\ and\
  \citenamefont {Bloch}}]{atala2013direct}%
  \BibitemOpen
  \bibfield  {author} {\bibinfo {author} {\bibfnamefont {M.}~\bibnamefont
  {Atala}}, \bibinfo {author} {\bibfnamefont {M.}~\bibnamefont {Aidelsburger}},
  \bibinfo {author} {\bibfnamefont {J.~T.}\ \bibnamefont {Barreiro}}, \bibinfo
  {author} {\bibfnamefont {D.}~\bibnamefont {Abanin}}, \bibinfo {author}
  {\bibfnamefont {T.}~\bibnamefont {Kitagawa}}, \bibinfo {author}
  {\bibfnamefont {E.}~\bibnamefont {Demler}}, \ and\ \bibinfo {author}
  {\bibfnamefont {I.}~\bibnamefont {Bloch}},\ }\href@noop {} {\bibfield
  {journal} {\bibinfo  {journal} {Nature {P}hysics}\ }\textbf {\bibinfo
  {volume} {9}},\ \bibinfo {pages} {795} (\bibinfo {year} {2013})}\BibitemShut
  {NoStop}%
\bibitem [{\citenamefont {Aidelsburger}\ \emph {et~al.}(2015)\citenamefont
  {Aidelsburger}, \citenamefont {Lohse}, \citenamefont {Schweizer},
  \citenamefont {Atala}, \citenamefont {Barreiro}, \citenamefont {Nascimbene},
  \citenamefont {Cooper}, \citenamefont {Bloch},\ and\ \citenamefont
  {Goldman}}]{aidelsburger2015measuring}%
  \BibitemOpen
  \bibfield  {author} {\bibinfo {author} {\bibfnamefont {M.}~\bibnamefont
  {Aidelsburger}}, \bibinfo {author} {\bibfnamefont {M.}~\bibnamefont {Lohse}},
  \bibinfo {author} {\bibfnamefont {C.}~\bibnamefont {Schweizer}}, \bibinfo
  {author} {\bibfnamefont {M.}~\bibnamefont {Atala}}, \bibinfo {author}
  {\bibfnamefont {J.~T.}\ \bibnamefont {Barreiro}}, \bibinfo {author}
  {\bibfnamefont {S.}~\bibnamefont {Nascimbene}}, \bibinfo {author}
  {\bibfnamefont {N.}~\bibnamefont {Cooper}}, \bibinfo {author} {\bibfnamefont
  {I.}~\bibnamefont {Bloch}}, \ and\ \bibinfo {author} {\bibfnamefont
  {N.}~\bibnamefont {Goldman}},\ }\href@noop {} {\bibfield  {journal} {\bibinfo
   {journal} {Nature {P}hysics}\ }\textbf {\bibinfo {volume} {11}},\ \bibinfo
  {pages} {162} (\bibinfo {year} {2015})}\BibitemShut {NoStop}%
\bibitem [{\citenamefont {Zeuner}\ \emph {et~al.}(2015)\citenamefont {Zeuner},
  \citenamefont {Rechtsman}, \citenamefont {Plotnik}, \citenamefont {Lumer},
  \citenamefont {Nolte}, \citenamefont {Rudner}, \citenamefont {Segev},\ and\
  \citenamefont {Szameit}}]{zeuner2015observation}%
  \BibitemOpen
  \bibfield  {author} {\bibinfo {author} {\bibfnamefont {J.~M.}\ \bibnamefont
  {Zeuner}}, \bibinfo {author} {\bibfnamefont {M.~C.}\ \bibnamefont
  {Rechtsman}}, \bibinfo {author} {\bibfnamefont {Y.}~\bibnamefont {Plotnik}},
  \bibinfo {author} {\bibfnamefont {Y.}~\bibnamefont {Lumer}}, \bibinfo
  {author} {\bibfnamefont {S.}~\bibnamefont {Nolte}}, \bibinfo {author}
  {\bibfnamefont {M.~S.}\ \bibnamefont {Rudner}}, \bibinfo {author}
  {\bibfnamefont {M.}~\bibnamefont {Segev}}, \ and\ \bibinfo {author}
  {\bibfnamefont {A.}~\bibnamefont {Szameit}},\ }\href@noop {} {\bibfield
  {journal} {\bibinfo  {journal} {Physical {R}eview {L}etters}\ }\textbf
  {\bibinfo {volume} {115}},\ \bibinfo {pages} {040402} (\bibinfo {year}
  {2015})}\BibitemShut {NoStop}%
\bibitem [{\citenamefont {Lumer}\ \emph {et~al.}(2013)\citenamefont {Lumer},
  \citenamefont {Plotnik}, \citenamefont {Rechtsman},\ and\ \citenamefont
  {Segev}}]{lumer2013self}%
  \BibitemOpen
  \bibfield  {author} {\bibinfo {author} {\bibfnamefont {Y.}~\bibnamefont
  {Lumer}}, \bibinfo {author} {\bibfnamefont {Y.}~\bibnamefont {Plotnik}},
  \bibinfo {author} {\bibfnamefont {M.~C.}\ \bibnamefont {Rechtsman}}, \ and\
  \bibinfo {author} {\bibfnamefont {M.}~\bibnamefont {Segev}},\ }\href@noop {}
  {\bibfield  {journal} {\bibinfo  {journal} {{P}hysical {R}eview {L}etters}\
  }\textbf {\bibinfo {volume} {111}},\ \bibinfo {pages} {243905} (\bibinfo
  {year} {2013})}\BibitemShut {NoStop}%
\bibitem [{\citenamefont {Ablowitz}\ \emph {et~al.}(2014)\citenamefont
  {Ablowitz}, \citenamefont {Curtis},\ and\ \citenamefont
  {Ma}}]{ablowitz2014linear}%
  \BibitemOpen
  \bibfield  {author} {\bibinfo {author} {\bibfnamefont {M.~J.}\ \bibnamefont
  {Ablowitz}}, \bibinfo {author} {\bibfnamefont {C.~W.}\ \bibnamefont
  {Curtis}}, \ and\ \bibinfo {author} {\bibfnamefont {Y.-P.}\ \bibnamefont
  {Ma}},\ }\href@noop {} {\bibfield  {journal} {\bibinfo  {journal} {Physical
  Review A}\ }\textbf {\bibinfo {volume} {90}},\ \bibinfo {pages} {023813}
  (\bibinfo {year} {2014})}\BibitemShut {NoStop}%
\bibitem [{\citenamefont {Katan}\ \emph {et~al.}(2016)\citenamefont {Katan},
  \citenamefont {Bekenestein}, \citenamefont {Bandres}, \citenamefont {Lumer},
  \citenamefont {Yonatan},\ and\ \citenamefont {Segev}}]{katan2016induction}%
  \BibitemOpen
  \bibfield  {author} {\bibinfo {author} {\bibfnamefont {Y.~T.}\ \bibnamefont
  {Katan}}, \bibinfo {author} {\bibfnamefont {R.}~\bibnamefont {Bekenestein}},
  \bibinfo {author} {\bibfnamefont {M.~A.}\ \bibnamefont {Bandres}}, \bibinfo
  {author} {\bibfnamefont {Y.}~\bibnamefont {Lumer}}, \bibinfo {author}
  {\bibfnamefont {P.}~\bibnamefont {Yonatan}}, \ and\ \bibinfo {author}
  {\bibfnamefont {M.}~\bibnamefont {Segev}},\ }in\ \href@noop {} {\emph
  {\bibinfo {booktitle} {CLEO: QELS\_Fundamental Science}}}\ (\bibinfo
  {organization} {Optical Society of America},\ \bibinfo {year} {2016})\ pp.\
  \bibinfo {pages} {FM3A--6}\BibitemShut {NoStop}%
\bibitem [{\citenamefont {Leykam}\ and\ \citenamefont
  {Chong}(2016)}]{leykam2016edge}%
  \BibitemOpen
  \bibfield  {author} {\bibinfo {author} {\bibfnamefont {D.}~\bibnamefont
  {Leykam}}\ and\ \bibinfo {author} {\bibfnamefont {Y.~D.}\ \bibnamefont
  {Chong}},\ }\href@noop {} {\bibfield  {journal} {\bibinfo  {journal}
  {{P}hysical {R}eview {L}etters}\ }\textbf {\bibinfo {volume} {117}},\
  \bibinfo {pages} {143901} (\bibinfo {year} {2016})}\BibitemShut {NoStop}%
\bibitem [{\citenamefont {Leykam}\ \emph {et~al.}(2018)\citenamefont {Leykam},
  \citenamefont {Mittal}, \citenamefont {Hafezi},\ and\ \citenamefont
  {Chong}}]{leykam2018reconfigurable}%
  \BibitemOpen
  \bibfield  {author} {\bibinfo {author} {\bibfnamefont {D.}~\bibnamefont
  {Leykam}}, \bibinfo {author} {\bibfnamefont {S.}~\bibnamefont {Mittal}},
  \bibinfo {author} {\bibfnamefont {M.}~\bibnamefont {Hafezi}}, \ and\ \bibinfo
  {author} {\bibfnamefont {Y.~D.}\ \bibnamefont {Chong}},\ }\href@noop {}
  {\bibfield  {journal} {\bibinfo  {journal} {Physical {R}eview {L}etters}\
  }\textbf {\bibinfo {volume} {121}},\ \bibinfo {pages} {023901} (\bibinfo
  {year} {2018})}\BibitemShut {NoStop}%
\bibitem [{\citenamefont {Bukov}\ \emph {et~al.}(2015)\citenamefont {Bukov},
  \citenamefont {D'Alessio},\ and\ \citenamefont
  {Polkovnikov}}]{bukov2015universal}%
  \BibitemOpen
  \bibfield  {author} {\bibinfo {author} {\bibfnamefont {M.}~\bibnamefont
  {Bukov}}, \bibinfo {author} {\bibfnamefont {L.}~\bibnamefont {D'Alessio}}, \
  and\ \bibinfo {author} {\bibfnamefont {A.}~\bibnamefont {Polkovnikov}},\
  }\href@noop {} {\bibfield  {journal} {\bibinfo  {journal} {Advances in
  {P}hysics}\ }\textbf {\bibinfo {volume} {64}},\ \bibinfo {pages} {139}
  (\bibinfo {year} {2015})}\BibitemShut {NoStop}%
\bibitem [{\citenamefont {Haldane}(1988)}]{haldane1988model}%
  \BibitemOpen
  \bibfield  {author} {\bibinfo {author} {\bibfnamefont {F.~D.~M.}\
  \bibnamefont {Haldane}},\ }\href {\doibase 10.1103/PhysRevLett.61.2015}
  {\bibfield  {journal} {\bibinfo  {journal} {Phys Rev Lett}\ }\textbf
  {\bibinfo {volume} {61}},\ \bibinfo {pages} {2015} (\bibinfo {year}
  {1988})}\BibitemShut {NoStop}%
\bibitem [{\citenamefont {Malomed}\ and\ \citenamefont
  {Kevrekidis}(2001)}]{malomed2001discrete}%
  \BibitemOpen
  \bibfield  {author} {\bibinfo {author} {\bibfnamefont {B.}~\bibnamefont
  {Malomed}}\ and\ \bibinfo {author} {\bibfnamefont {P.}~\bibnamefont
  {Kevrekidis}},\ }\href@noop {} {\bibfield  {journal} {\bibinfo  {journal}
  {Physical Review E}\ }\textbf {\bibinfo {volume} {64}},\ \bibinfo {pages}
  {026601} (\bibinfo {year} {2001})}\BibitemShut {NoStop}%
\bibitem [{\citenamefont {Kevrekidis}\ \emph {et~al.}(2001)\citenamefont
  {Kevrekidis}, \citenamefont {Malomed},\ and\ \citenamefont
  {Bishop}}]{kevrekidis2001bound}%
  \BibitemOpen
  \bibfield  {author} {\bibinfo {author} {\bibfnamefont {P.}~\bibnamefont
  {Kevrekidis}}, \bibinfo {author} {\bibfnamefont {B.}~\bibnamefont {Malomed}},
  \ and\ \bibinfo {author} {\bibfnamefont {A.}~\bibnamefont {Bishop}},\
  }\href@noop {} {\bibfield  {journal} {\bibinfo  {journal} {Journal of Physics
  A: Mathematical and General}\ }\textbf {\bibinfo {volume} {34}},\ \bibinfo
  {pages} {9615} (\bibinfo {year} {2001})}\BibitemShut {NoStop}%
\bibitem [{\citenamefont {Christodoulides}\ and\ \citenamefont
  {Joseph}(1988)}]{christodoulides1988discrete}%
  \BibitemOpen
  \bibfield  {author} {\bibinfo {author} {\bibfnamefont {D.}~\bibnamefont
  {Christodoulides}}\ and\ \bibinfo {author} {\bibfnamefont {R.}~\bibnamefont
  {Joseph}},\ }\href@noop {} {\bibfield  {journal} {\bibinfo  {journal}
  {{O}ptics {L}etters}\ }\textbf {\bibinfo {volume} {13}},\ \bibinfo {pages}
  {794} (\bibinfo {year} {1988})}\BibitemShut {NoStop}%
\bibitem [{\citenamefont {Eggleton}\ \emph {et~al.}(1996)\citenamefont
  {Eggleton}, \citenamefont {Slusher}, \citenamefont {de~Sterke}, \citenamefont
  {Krug},\ and\ \citenamefont {Sipe}}]{eggleton1996bragg}%
  \BibitemOpen
  \bibfield  {author} {\bibinfo {author} {\bibfnamefont {B.~J.}\ \bibnamefont
  {Eggleton}}, \bibinfo {author} {\bibfnamefont {R.}~\bibnamefont {Slusher}},
  \bibinfo {author} {\bibfnamefont {C.~M.}\ \bibnamefont {de~Sterke}}, \bibinfo
  {author} {\bibfnamefont {P.~A.}\ \bibnamefont {Krug}}, \ and\ \bibinfo
  {author} {\bibfnamefont {J.}~\bibnamefont {Sipe}},\ }\href@noop {} {\bibfield
   {journal} {\bibinfo  {journal} {{P}hysical {R}eview {L}etters}\ }\textbf
  {\bibinfo {volume} {76}},\ \bibinfo {pages} {1627} (\bibinfo {year}
  {1996})}\BibitemShut {NoStop}%
\bibitem [{\citenamefont {Stegeman}\ and\ \citenamefont
  {Segev}(1999)}]{stegeman1999optical}%
  \BibitemOpen
  \bibfield  {author} {\bibinfo {author} {\bibfnamefont {G.~I.}\ \bibnamefont
  {Stegeman}}\ and\ \bibinfo {author} {\bibfnamefont {M.}~\bibnamefont
  {Segev}},\ }\href@noop {} {\bibfield  {journal} {\bibinfo  {journal}
  {Science}\ }\textbf {\bibinfo {volume} {286}},\ \bibinfo {pages} {1518}
  (\bibinfo {year} {1999})}\BibitemShut {NoStop}%
\bibitem [{\citenamefont {Kivshar}\ and\ \citenamefont
  {Agrawal}(2003)}]{kivshar2003optical}%
  \BibitemOpen
  \bibfield  {author} {\bibinfo {author} {\bibfnamefont {Y.~S.}\ \bibnamefont
  {Kivshar}}\ and\ \bibinfo {author} {\bibfnamefont {G.}~\bibnamefont
  {Agrawal}},\ }\href@noop {} {\emph {\bibinfo {title} {Optical solitons: from
  fibers to photonic crystals}}}\ (\bibinfo  {publisher} {Academic press},\
  \bibinfo {year} {2003})\BibitemShut {NoStop}%
\bibitem [{\citenamefont {Fleischer}\ \emph {et~al.}(2003)\citenamefont
  {Fleischer}, \citenamefont {Segev}, \citenamefont {Efremidis},\ and\
  \citenamefont {Christodoulides}}]{fleischer2003observation}%
  \BibitemOpen
  \bibfield  {author} {\bibinfo {author} {\bibfnamefont {J.~W.}\ \bibnamefont
  {Fleischer}}, \bibinfo {author} {\bibfnamefont {M.}~\bibnamefont {Segev}},
  \bibinfo {author} {\bibfnamefont {N.~K.}\ \bibnamefont {Efremidis}}, \ and\
  \bibinfo {author} {\bibfnamefont {D.~N.}\ \bibnamefont {Christodoulides}},\
  }\href@noop {} {\bibfield  {journal} {\bibinfo  {journal} {Nature}\ }\textbf
  {\bibinfo {volume} {422}},\ \bibinfo {pages} {147} (\bibinfo {year}
  {2003})}\BibitemShut {NoStop}%
\bibitem [{\citenamefont {Weinstein}(1999)}]{weinstein1999excitation}%
  \BibitemOpen
  \bibfield  {author} {\bibinfo {author} {\bibfnamefont {M.~I.}\ \bibnamefont
  {Weinstein}},\ }\href@noop {} {\bibfield  {journal} {\bibinfo  {journal}
  {Nonlinearity}\ }\textbf {\bibinfo {volume} {12}},\ \bibinfo {pages} {673}
  (\bibinfo {year} {1999})}\BibitemShut {NoStop}%
\bibitem [{\citenamefont {Merle}(1988)}]{merle1988existence}%
  \BibitemOpen
  \bibfield  {author} {\bibinfo {author} {\bibfnamefont {F.}~\bibnamefont
  {Merle}},\ }\href@noop {} {\bibfield  {journal} {\bibinfo  {journal} {Journal
  of differential equations}\ }\textbf {\bibinfo {volume} {74}},\ \bibinfo
  {pages} {50} (\bibinfo {year} {1988})}\BibitemShut {NoStop}%
\bibitem [{\citenamefont {Esteban}\ and\ \citenamefont
  {S{\'e}r{\'e}}(1995)}]{esteban1995stationary}%
  \BibitemOpen
  \bibfield  {author} {\bibinfo {author} {\bibfnamefont {M.~J.}\ \bibnamefont
  {Esteban}}\ and\ \bibinfo {author} {\bibfnamefont {{\'E}.}~\bibnamefont
  {S{\'e}r{\'e}}},\ }\href@noop {} {\bibfield  {journal} {\bibinfo  {journal}
  {Communications in {M}athematical {P}hysics}\ }\textbf {\bibinfo {volume}
  {171}},\ \bibinfo {pages} {323} (\bibinfo {year} {1995})}\BibitemShut
  {NoStop}%
\bibitem [{\citenamefont {Esteban}\ \emph {et~al.}(1996)\citenamefont
  {Esteban}, \citenamefont {Georgiev},\ and\ \citenamefont
  {S{\'e}r{\'e}}}]{esteban1996stationary}%
  \BibitemOpen
  \bibfield  {author} {\bibinfo {author} {\bibfnamefont {M.~J.}\ \bibnamefont
  {Esteban}}, \bibinfo {author} {\bibfnamefont {V.}~\bibnamefont {Georgiev}}, \
  and\ \bibinfo {author} {\bibfnamefont {E.}~\bibnamefont {S{\'e}r{\'e}}},\
  }\href@noop {} {\bibfield  {journal} {\bibinfo  {journal} {Calculus of
  Variations and Partial Differential Equations}\ }\textbf {\bibinfo {volume}
  {4}},\ \bibinfo {pages} {265} (\bibinfo {year} {1996})}\BibitemShut {NoStop}%
\bibitem [{\citenamefont {Esteban}\ and\ \citenamefont
  {S{\'e}r{\'e}}(2002)}]{esteban2002overview}%
  \BibitemOpen
  \bibfield  {author} {\bibinfo {author} {\bibfnamefont {M.~J.}\ \bibnamefont
  {Esteban}}\ and\ \bibinfo {author} {\bibfnamefont {E.}~\bibnamefont
  {S{\'e}r{\'e}}},\ }\href@noop {} {\bibfield  {journal} {\bibinfo  {journal}
  {Discrete \& {C}ontinuous {D}ynamical Systems-A}\ }\textbf {\bibinfo {volume}
  {8}},\ \bibinfo {pages} {381} (\bibinfo {year} {2002})}\BibitemShut {NoStop}%
\bibitem [{\citenamefont {Boussaid}\ and\ \citenamefont
  {Comech}(2017)}]{boussaid2017nonrelativistic}%
  \BibitemOpen
  \bibfield  {author} {\bibinfo {author} {\bibfnamefont {N.}~\bibnamefont
  {Boussaid}}\ and\ \bibinfo {author} {\bibfnamefont {A.}~\bibnamefont
  {Comech}},\ }\href@noop {} {\bibfield  {journal} {\bibinfo  {journal} {SIAM
  Journal on Mathematical Analysis}\ }\textbf {\bibinfo {volume} {49}},\
  \bibinfo {pages} {2527} (\bibinfo {year} {2017})}\BibitemShut {NoStop}%
\bibitem [{\citenamefont {Karplus}\ and\ \citenamefont
  {Luttinger}(1954)}]{karplus1954hall}%
  \BibitemOpen
  \bibfield  {author} {\bibinfo {author} {\bibfnamefont {R.}~\bibnamefont
  {Karplus}}\ and\ \bibinfo {author} {\bibfnamefont {J.}~\bibnamefont
  {Luttinger}},\ }\href@noop {} {\bibfield  {journal} {\bibinfo  {journal}
  {Physical {R}eview}\ }\textbf {\bibinfo {volume} {95}},\ \bibinfo {pages}
  {1154} (\bibinfo {year} {1954})}\BibitemShut {NoStop}%
\bibitem [{\citenamefont {Dreisow}\ \emph {et~al.}(2010)\citenamefont
  {Dreisow}, \citenamefont {Heinrich}, \citenamefont {Keil}, \citenamefont
  {T{\"u}nnermann}, \citenamefont {Nolte}, \citenamefont {Longhi},\ and\
  \citenamefont {Szameit}}]{dreisow2010classical}%
  \BibitemOpen
  \bibfield  {author} {\bibinfo {author} {\bibfnamefont {F.}~\bibnamefont
  {Dreisow}}, \bibinfo {author} {\bibfnamefont {M.}~\bibnamefont {Heinrich}},
  \bibinfo {author} {\bibfnamefont {R.}~\bibnamefont {Keil}}, \bibinfo {author}
  {\bibfnamefont {A.}~\bibnamefont {T{\"u}nnermann}}, \bibinfo {author}
  {\bibfnamefont {S.}~\bibnamefont {Nolte}}, \bibinfo {author} {\bibfnamefont
  {S.}~\bibnamefont {Longhi}}, \ and\ \bibinfo {author} {\bibfnamefont
  {A.}~\bibnamefont {Szameit}},\ }\href@noop {} {\bibfield  {journal} {\bibinfo
   {journal} {{P}hysical {R}eview {L}etters}\ }\textbf {\bibinfo {volume}
  {105}},\ \bibinfo {pages} {143902} (\bibinfo {year} {2010})}\BibitemShut
  {NoStop}%
\end{thebibliography}%

\end{document}